\documentclass[aps,prb,preprint,superscriptaddress,onecolumn,showpacs,amsmath,floatfix,citeautoscript]{revtex4}

\usepackage[]{amsmath}
\usepackage{amssymb}
\usepackage[dvips]{graphicx}
\usepackage{color}
\usepackage{bm}
\usepackage{tabularx}
\usepackage{mathtools}
\usepackage{algpseudocode}
\usepackage{enumitem}
\usepackage{multirow}
\usepackage{epstopdf}
\usepackage{ulem}

\newcommand{\rom}[1]{\uppercase\expandafter{\romannumeral #1\relax}}

\newcommand{\norm}[2]{\lVert {#1} \rVert_{#2}}

\newcommand{\xie}{\mathcal}

\newcommand{\yu}{\mathbb}

\begin{document}
\title{Warm dense matter simulation via electron temperature dependent deep potential molecular dynamics}

\author{Yuzhi Zhang}
\affiliation{Center for Applied Physics and Technology, HEDPS, College of Engineering, Peking University, Beijing 100871, People's Republic of China}
\affiliation{Beijing Institute of Big Data Research, Beijing 100871, People's Republic of China}
\author{Chang Gao}
\affiliation{Center for Applied Physics and Technology, HEDPS, College of Engineering, Peking University, Beijing 100871, People's Republic of China}
\author{Linfeng Zhang}
\email{linfengz@princeton.edu}
\affiliation{Program in Applied and Computational Mathematics, Princeton University, Princeton, NJ, USA}
\author{Han Wang}
\email{wang_han@iapcm.ac.cn}
\affiliation{Laboratory of Computational Physics, Institute of Applied Physics and Computational Mathematics, Huayuan Road 6, Beijing 100088, People's Republic of China}
  \author{Mohan Chen}
  \email{mohanchen@pku.edu.cn}
\affiliation{Center for Applied Physics and Technology, HEDPS, College of Engineering, Peking University, Beijing 100871, People's Republic of China}

\date{\today}

\begin{abstract}
Simulating warm dense matter that undergoes a wide range of temperatures and densities is challenging.
Predictive theoretical models, such as quantum-mechanics-based first-principles molecular dynamics (FPMD), require a huge amount of computational resources.
Herein, we propose a deep learning based scheme, called electron temperature dependent deep potential molecular dynamics (TDDPMD),
for efficiently simulating warm dense matter with the accuracy of FPMD.
The TDDPMD simulation is several orders of magnitudes faster than FPMD, and, unlike FPMD,
its efficiency is not affected by the electron temperature.
We apply the TDDPMD scheme to beryllium (Be) in a wide range of temperatures (0.4 to 2500 eV)
and densities (3.50 to 8.25 g/cm$^3$).
Our results demonstrate that the TDDPMD method not only accurately reproduces the structural properties of
Be along the principal Hugoniot curve at the FPMD level, but also yields even more reliable diffusion
coefficients than typical FPMD simulations due to its ability to simulate larger systems with longer time.
\end{abstract}

\maketitle
\section{INTRODUCTION}
Materials under extreme conditions exhibit rich physics.
Of particular interest is the warm dense matter~\cite{14book-wdm}, which is formed by partially ionized electrons that interact strongly with the nuclei, 
and has recently attracted much attention due to its vital role in astrophysics~\cite{06RMP-Remington} and inertial confinement fusion~\cite{98PP-Bodner,95PP-Lindl,16PP-He}.
However, the corresponding experimental data are largely limited due to enormous challenges in conducting experiments in such conditions.
This also sets high obstacles to developing empirical models and theories.
Therefore, it is typically a necessity to adopt quantum-mechanics-based first-principles models to describe the strong couplings between electrons and ions at finite temperatures.
For example, widely adopted methods include first-principles molecular dynamics (FPMD)~\cite{13E-Wang,13L-White,16PP-Zhang,18L-Mo,19B-Zhang} 
based on the density functional theory (DFT) \cite{64PR-Hohenberg,65PR-Kohn,65PR-Mermin} and the path-integral Monte Carlo (PIMC) method~\cite{95RMP-Ceperley,10L-Hu,11B-Hu,13L-Brown,15L-Militzer}.
Although these methods provide by far the most accurate descriptions for warm dense matter, the quantum mechanics algorithms adopted in these methods are computationally very expensive.

In this context, it has been a long-standing goal to develop quantum-mechanics based models that are computationally efficient.
The DFT method can be categorized into the
Kohn-Sham framework (KSDFT)~\cite{65PR-Kohn} and the orbital-free scheme (OFDFT) ~\cite{12CPC-Karasiev,18JMR-Witt}.
The KSDFT framework that explicitly includes single-particle orbitals is capable of describing electrons in both free and bound states.
Nevertheless, the unfavorable cubic scaling of KSDFT, when combined with the molecular dynamics method, limits the system size and trajectory length up to a few tens of atoms and picoseconds, respectively.
Worse still, the computational cost of KSDFT becomes a more severe problem when the electron temperature increases,
as more Kohn-Sham orbitals at higher energies are required to represent the Fermi-Dirac distribution of electrons with sufficient accuracy.
Meanwhile, a hard pseudopotential with a high energy cutoff is unavoidably needed to accurately characterize ion-electron interactions at elevated temperatures.
These issues lead to large error bars and artificial size effects~\cite{04JPCB-Yeh} in evaluating important properties of warm dense matter such as the diffusion coefficient.
Moreover, the traditional KSDFT can only simulate systems at electron temperatures up to the order of 10 eV \cite{01L-Surh,10L-Hu,13PP-Wang,14L-Sjo,14E-She}.
This situation was alleviated by an extended KSDFT scheme~\cite{16PP-Zhang}, which treats the high-energy electronic states analytically by using the plane-wave basis sets,
but the intrinsic cubic scaling of KSDFT still remains a problem in the extended KSDFT scheme.
On the other hand, 
the OFDFT scheme is relatively more efficient than KSDFT because
the former one is suitable for describing free-electron-like systems, and has been used in studying warm dense matter~\cite{13L-White,14L-Sjo}.
However, it was demonstrated that OFDFT is not adequate to characterize electrons in partially ionized shells~\cite{16B-Gao}. 
Besides the DFT methods, the PIMC method is suitable to study materials at extremely high
temperatures, but also faces a severe problem of efficiency at relatively lower temperatures~\cite{95RMP-Ceperley}.
Notably, PIMC cannot yield time-dependent transport properties such as the diffusion coefficients.

Recently proposed machine learning based approaches have shown promising potentials in reproducing FPMD accuracy
with a dramatic enhancement of the efficiency~\cite{behler2007generalized,chmiela2017machine,
smith2017ani,schutt2017quantum,han2017deep,18L-Zhang}. 
Of particular relevance to this work is the deep potential molecular dynamics (DPMD) model~\cite{18L-Zhang,18CPC-Wang},
which, upon training with {\it ab initio} data, is capable of generating a many-body potential energy surface
and sampling much larger size and time scales without loss of accuracy.
In this regard, the DPMD method plays an important role in reducing statistical errors and size effects of those computed properties of materials.
However, there are two issues in directly applying DPMD to study warm dense matter.
First, while DPMD parameterizes a potential energy surface, our goal here is to accurately represent a free energy surface, which depends not only on atomic positions and their chemical species, but also on a wide range of electron temperature $T$.
Therefore, we need to suitably extend the current machinery for the potential energy surface. 
Second, the magnitude of the fluctuations of the free energy surface and the atomic forces for warm dense matter changes drastically with electron temperature,
causing difficulties in finding a suitable DPMD model to adequately describe this system.

In this work, we propose a temperature-dependent DPMD (TDDPMD) method, which inherits from the DPMD model the essential physical considerations and characterizes the relation between the free energy surface and electron temperatures $T$.
Compared with the typical FPMD method that has a cubic scaling, the TDDPMD is a linearly scaling method whose efficiency does not rely on the electron temperature.
Here we take warm dense beryllium (Be) as an example to compute its equation of state, which has been studied in previous works~\cite{13E-Wang,15PP-Li,18arxiv-Gao}.
Upon training with first-principles data, we utilize the TDDPMD model to
evaluate the principal Hugoniot curve of Be
with its ion densities ranging from 3.50 to 8.25 g/cm$^3$, and electron temperatures ranging from 0.4 to 2500 eV.
The two Hugoniot curves from FPMD and TDDPMD agree very well.
Additionally, more structural and dynamical properties of Be are computed and analyzed by using the TDDPMD simulations.
Our work demonstrates that the TDDPMD scheme owns the accuracy of quantum-mechanics-based first-principles methods and is efficient for studying warm dense matter via large systems and long trajectories. We expect TDDPMD to have a profound impact in improving our understanding of warm dense matter and other materials in extreme conditions.

\section{Methods}

Consider a system composed of $N$ atoms and $N_e$ electrons.
The Cartesian coordinates of the $N$ atoms are denoted by
$\bm{R} = \{\bm{R}_1, \cdots,  \bm{R}_I, \cdots, \bm{R}_N\}$.
The quantity of our interest is a free energy surface $E(\bm{R},T)$,
where $T$ depicts the electron temperature.
$E(\bm{R},T)$ is obtained by minimizing the Mermin free energy according to the finite temperature DFT\cite{65PR-Mermin}.
In practical DFT simulations, the free energy can also be expressed as
\begin{equation}\label{fes}
    E[\{\psi_i\},\{f_i\}]=U[\{\psi_i\},\{f_i\}]-T \cdot S[\{f_i\}],
\end{equation}
where $U$ is the internal energy and $S$ is the entropy.
Additionally, the Kohn-Sham orbitals $\{\psi_i\}$
meet the requirement of  $\label{psi}
\int\psi_i^*(\mathbf{r})\psi_j(\mathbf{r})d\mathbf{r}=\delta_{ij}$
and the fractional occupation $\{f_{i}\}$ are determined by the Fermi-Dirac distribution of electrons.

Similar to the philosophy adopted for constructing a typical potential energy surface model~\cite{18L-Zhang},
keys to our considerations are some physical constraints:
1) the extensive property of $E$, i.e., $E$ should scale proportionally with the atomic number $N$;
and 2) the symmetric property of $E$, i.e., $E$ should remain invariant upon translation, rotation,
and identical particle permutation operations of the atomic positions.
Moreover, beyond a typical potential energy surface model, we need to introduce explicit dependence of $E$ on the electron temperature $T$.
Due to the wide range of electron temperatures considered in warm dense matter simulations,
the fluctuations of the free energies and the forces differ by several order of magnitudes,
resulting in an ill-posed fitting problem at its first appearance.
This issue will be addressed by incorporating in $E$
the scale dependence of these fluctuations on electron temperatures.
Finally, the atomic forces $\bm{F}_I=-\frac{\partial{E}}{\partial{\bm{R}_I}}$ and the stress tensor
$\bm{\Xi} = \frac{1}{V} \frac{\partial E}{\partial \bm{\Omega}}\cdot \bm{\Omega}^T$ are obtained analytically via the Hellman-Feynman theorem.
Here $\bm{\Omega}$ denotes the cell tensor and $\bm{\Omega}^T$ is its transpose, and $V=\det\bm{\Omega}$ is the cell volume.

\begin{figure}[htbp]
\begin{center}
\includegraphics[width = 8cm]{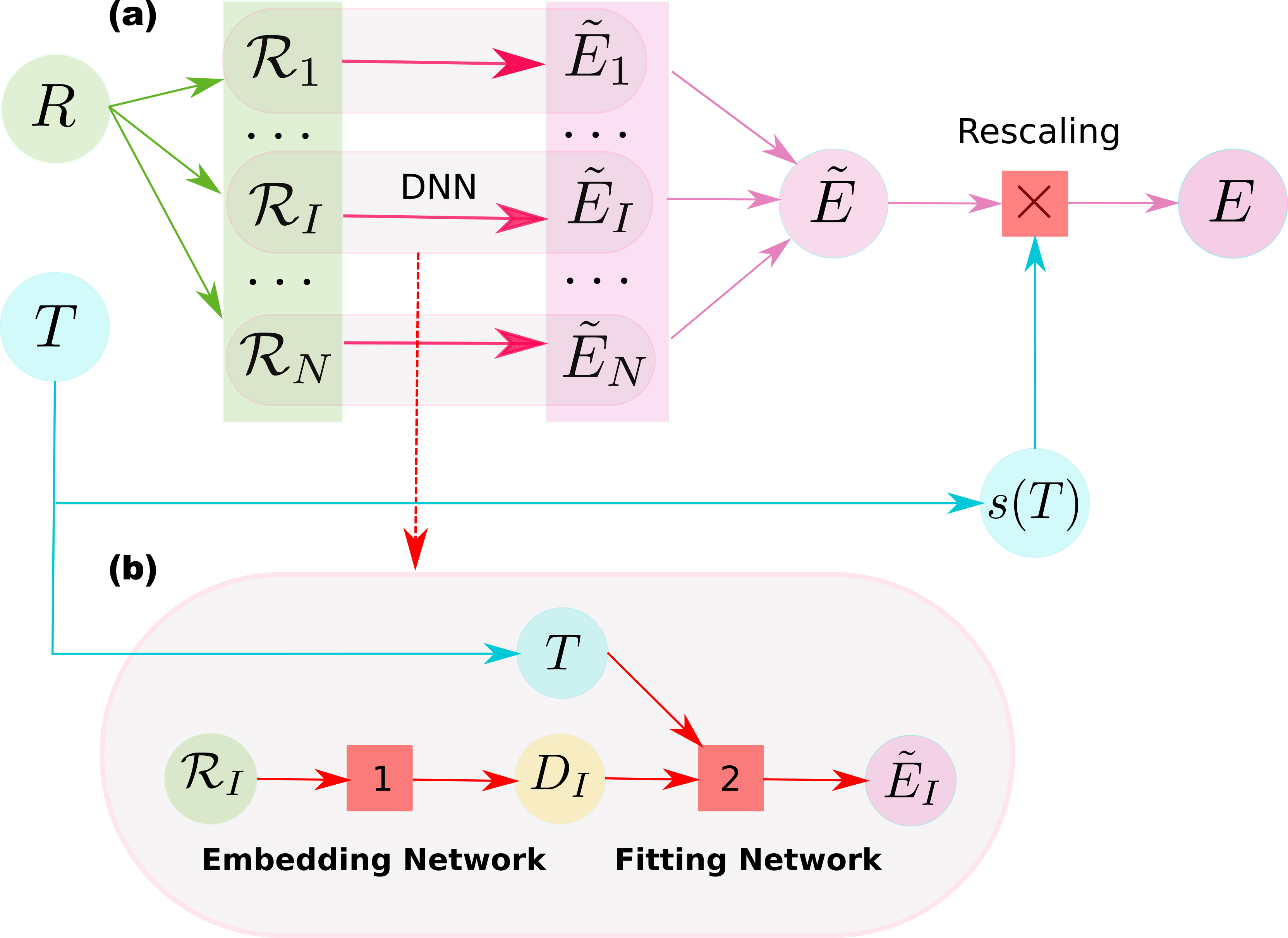}
\caption{(Color online) Schematic illustration of the temperature-dependent deep potential molecular dynamics (TPDPMD) method. (a) Mapping of the atomic coordinates $\bm{R} $ and electron temperature $T$ to the local free energy $\tilde{E}$. The rescaling step depicts a multiplication operator that links
$\tilde{E}$ and the scaling function $s(T)$ to the total free energy $E$. (b) Deep neural network (DNN) including the embedding network and the fitting network, which
are labelled as 1 and 2 in red rectangles, respectively.
The embedding network serves as a filter to extract information from local environment and outputs a descriptor $D_I$,
which includes information of atomic positions with preserved symmetries.
The fitting network maps $D_I$ and $T$ to a local scaled free energy $\tilde{E}_I$.
}
\label{fig:scheme}
\end{center}
\end{figure}

As shown in Eq.~\ref{model} and Fig.~\ref{fig:scheme} (a), we write $E(\bm{R},T)$ as the product of a temperature dependent scaling factor $s(T)$ and the sum of atomic contributions $\tilde E_I$:
\begin{equation}\label{model}
E(\bm{R},T) = s(T)\tilde E(\bm{R},T)=s(T)\sum_I \tilde E_I(\mathcal{R}_I,T)
\end{equation}
where $\mathcal{R}_I \in \yu{R}^{N_I \times 3} $ represents the local environment of atom $I$ as
$\mathcal{R}_I=\{\bm{r}^T_{1I},\cdots,
\bm{r}^T_{JI}, \cdots, \bm{r}^T_{N_I I} | J\in\mathcal{N}_{r_c}^I\}^T.$
Here we utilize the relative coordinates $\bm{r}_{JI}  \equiv \bm{r}_J - \bm{r}_I$ and set $r_{JI} = |\bm{r}_{JI} |$.
$\mathcal{N}_{r_c}^I$ depicts the index set of atoms $J$ neighboring $I$ within a cutoff radius $r_c$,  and $N_I = |\mathcal{N}_{r_c}^I|$ denotes its cardinality.
The construction of $\tilde{E}_I(\mathcal{R}_I,T)$ follows the smooth version idea of the deep potential model~\cite{18ANIPS-Zhang}, and is realized by the deep neural network.

\begin{figure}[htbp]
\begin{center}
\includegraphics[width = 8.4cm]{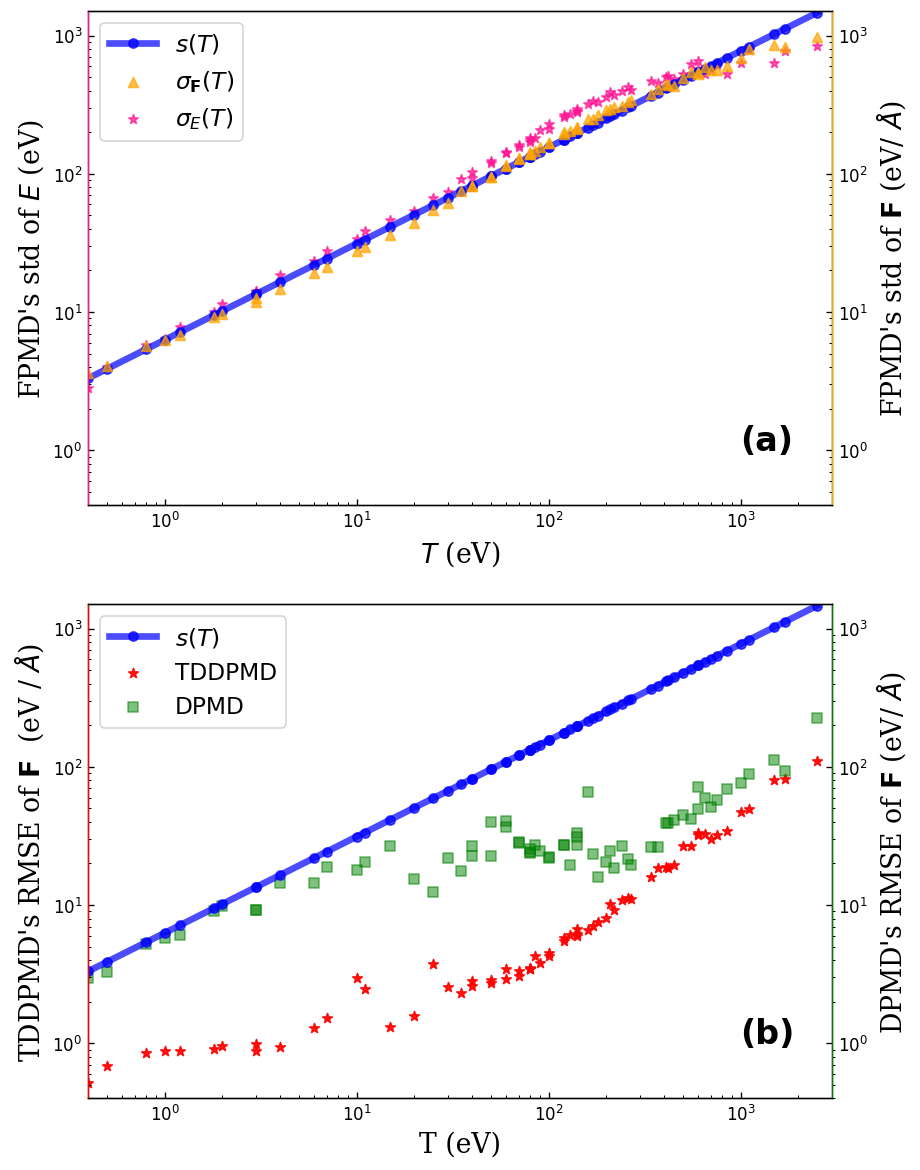}
\caption{(Color online) (a) Standard deviations of total free energy $\sigma(E)$ (in eV, pink stars) and
atomic forces $\sigma(F)$ (in eV/\AA, orange triangles). The fitted scaling factor $s(T)$
is shown with blue dotted lines, and is dimensionless. (b) RMSE on force predictions (in eV/\AA) of TDDPMD (red stars) and DPMD (green stars).  
}
\label{fig:fit}
\end{center}
\end{figure}

The scaling factor $s(T)$ is constructed in accordance with the dependence of the fluctuation of atomic forces on the electron temperature.
Taking the warm dense Be as an example,
in the training process, let $\sigma_{\bm{F}}(T)$ and $\sigma_{{E}}(T)$ denote the standard deviations of atomic forces $\bm{F}$ and free energies $E$ of a given system at temperature $T$, respectively.
According to Fig.~\ref{fig:fit}(a),
$\sigma_{\bm{F}}(T)$ ranges over more than two orders of magnitudes
when the simulated $T$ changes from 0.4 to 2500 eV
, and $\sigma_{E}(T)$ behaves similarly to $\sigma_{\bm{F}}(T)$.
Interestingly, $Log(\sigma_{\bm{F}}(T))$ depends approximately linearly on $Log(T)$.
Therefore, a least-square regression is used to estimate the linear coefficients $a$ and $b$, i.e.,
\begin{equation}
Log(\hat{\sigma}_{\bm{F}}(T)) = a \cdot Log(T) + b.
\end{equation}
$s(T)$ is then defined by the estimated fluctuation of forces:
\begin{equation}
s(T) \equiv \hat{\sigma}_{\bm{F}}(T) =  e^b \cdot T^a.
\end{equation}
Next, the network parameters are obtained by minimizing the loss function:
\begin{align}\nonumber
&\mathcal{L}(p_e, p_f, p_v) \\
&= \frac{1}{\vert \xie B\vert} \sum_{l \in \xie{B}}
\frac 1{s^2(T_l)}
\Bigr(p_e \Delta E_{l}^2 + p_f |\Delta \bm{F}_{l}|^2 + p_v \norm{\Delta \bm{\Xi}_l}{}^2 \Bigl),  \label{eq:loss}
\end{align}
where $p_e$, $p_f$, and $p_v$ are tunable prefactors, and $\xie{B}$ denotes the minibatch~\cite{14Adam} of training data with $l$ being the index.
As shown in Fig.~\ref{fig:fit}(b), the incorporation of the scaling factor $s(T)$ in the TDDPMD training process results in a much better-posed training process than the fitting with the DPMD scheme.
To be specific, given the same training data, the TDDPMD model with a scaling scheme produces much smaller root mean squared errors (RMSE) on free energy and atomic forces predictions than the traditional DMPD scheme that does not consider the scaling scheme, especially at lower temperatures.
%


We design the structure of the neural networks and the training scheme as follows.
The embedding network shown in Fig.~\ref{fig:scheme} is composed of three layers (25, 50, and 100 nodes)
while the fitting network has three hidden layers with 240 nodes in each layer.
The total number of training steps is set to 2,000,000,
with the size of minibatch being two.
The radius cutoff $r_c$ is chosen to be 5.0~\AA, and the inverse distance between atoms decays smoothly from 0.5 ~\AA\ to $r_c$.
The fitting parameters $p_e$, $p_f$, $p_v$ in Eq.~\ref{eq:loss}
are set to $(0.02, 1000, 0.02)$ at the beginning of training and gradually change to $(2.0, 1.0, 2.0)$.
%
In our TDDPMD simulations, we used cubic cells with the number of atoms ranging from 32 to 2048 atoms with periodic boundary conditions.
The NVT ensemble was adopted in Born-Oppenheimer molecular dynamics.
In order to simulate the Hugoniot curve, a time step of 10$^{-6}$ was adopted in TDDPMD for 1 million steps.
Furthermore, we ran TDDPMD for 32-, 256-, and 2048-atom cubic cells of Be to predict structural and dynamical properties in terms of different cell sizes, where time steps ranging from 10$^{-5}$ to 5$\times$10$^{-4}$ ps were chosen to run for 100 ps trajectories.
We used the FPMD data from Ref.~\onlinecite{18arxiv-Gao} as the training data.

The principal Hugoniot curve is obtained from the Rankine-Hugoniot equation as $U-U_{0}=\frac{1}{2}(P+P_{0})(V-V_{0}). $
Here $U$, $P$, and $V$ are the internal energy, pressure, and volume of a given system, respectively;
the three variables with a subscript of $0$ means a reference system at 300 K and 0 GPa.
The corresponding density is 1.84 g/cm$^3$.
We fit the entropy term $T \cdot S$ using the same methodology adopted for the free energy, and obtain internal energy by $U = E({\bf R}, T) + T \cdot S({\bf R}, T)$.

\section{Results and discussions}

As summarized in Tab.~\ref{tab}, we compare the efficiency of the FPMD and TDDPMD methods with respect to the number of atoms $N$ and temperature $T$.
The tests were run on a single CPU (Intel(R) Xeon(R) Gold 6126 CPU @ 2.60GHz) and the averaged time for one step molecular dynamics is listed.
In general, we find the TDDPMD method is much more efficient than the FPMD method in obtaining free energies and atomic forces from the trained deep neural network.
Specifically, by looking at the 32-atom systems at temperatures of 1, 70, and 2500 eV, we observe that the time for one step molecular dynamics in TDDPMD is a constant of around 0.36 second.
In stark contrast, the same operation costs 219 seconds at $T$=1 eV and 1122 seconds at $T$=2500 eV by using FPMD.
Moreover, it is observed that, for the systems of different sizes but the same density (8.1 g/cm$^3$) and temperature (70 eV), the advantage of TDDPMD over FPMD in terms of efficiency is more significant in larger systems.
This is due to the fact that TDDPMD scales linearly while the FPMD generally
has a cubic scaling with respect to the system size.
For example, TDDPMD is about 2.30$\times$10$^3$ and 6.28$\times$10$^4$ times
faster than FPMD in the 32- and 128-atom cells, respectively.

\begin{table}[htbp]
\caption{Simulation efficiency of FPMD and TDDPMD for Be in different conditions including number of atoms $N$, temperature $T$, and ion density $\rho$.
$t_D$ and $t_F$ denote the averaged wall time to run one step of molecular dynamics
by the TDDPMD and FPMD methods, respectively.
The results were obtained by performing a 100-step MD simulation for systems at each condition on Intel(R) Xeon(R) Gold 6126 CPU @ 2.60GHz, using a single CPU core.
}
\scalebox{1.10}{
\begin{tabular}{c|c|c|c|c|c}
  \hline
 $N$ & $T$ (eV) & $\rho (g/cm^3)$ & $t_D$ (s) & $t_F$ (s) &  $t_F/t_D$\\\hline
 32  & 1    & 4.0  & 0.36  & 219   & 6.08 $\times 10^2$  \\
 32  & 70   & 8.1  & 0.36  & 828   & 2.30$\times 10^3$ \\
 32  & 2500 & 7.45 & 0.34  & 1122  & 3.30 $\times 10^3$  \\
 64  & 70   & 8.1  & 0.68  & 9252  & 1.36$\times 10^4$ \\
 128 & 70   & 8.1  & 1.42 & 89172 & 6.28$\times 10^4$   \\
 256 & 70   & 8.1  & 2.80 & -    & -     \\
  \hline
\end{tabular}
\label{tab}
}
\end{table}

\begin{figure*}[htbp]
\begin{center}
\includegraphics[width = 16cm]{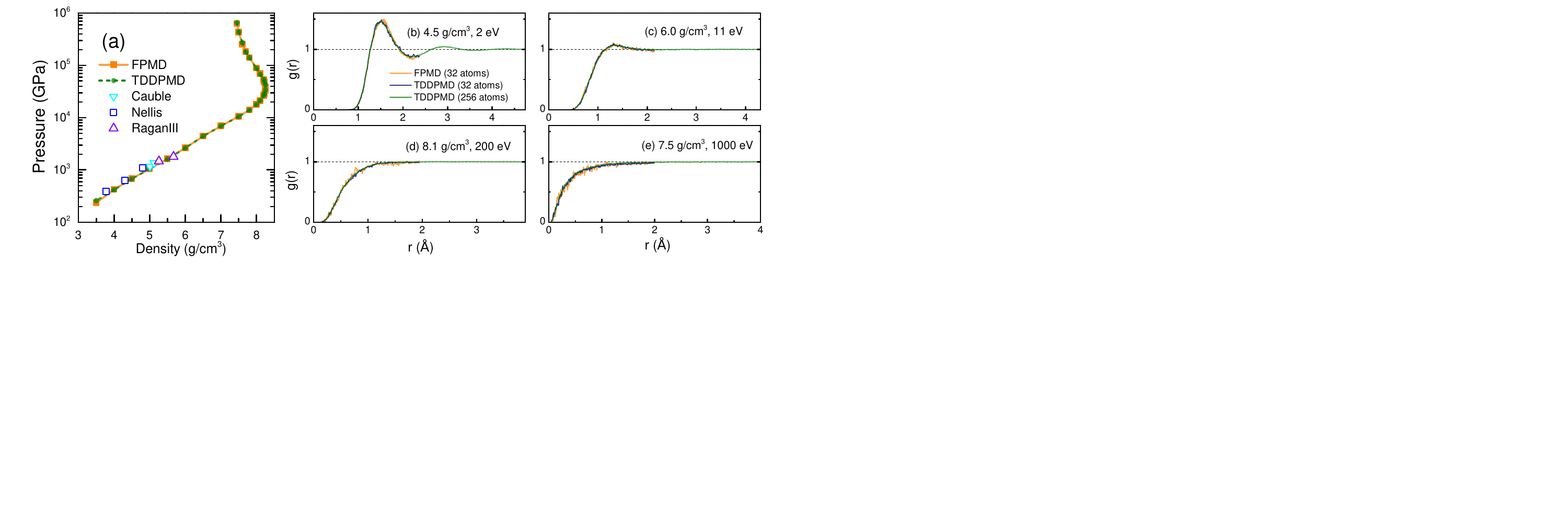}
\caption{(Color online) (a) Principal Hugoniot curves of Be obtained from the FPMD and TDDPMD simulations,
as well as the experimental data from Cauble {\it et al.} \cite{98L-Cauble},
Nellis {\it et al.} \cite{97JAP-Nellis},
and Ragan {\it et al.} \cite{82A-Ragan}.
(b-e) Radial distribution functions $g(r)$ of warm dense Be obtained from simulations. The densities and temperatures
are selected from the principal Hugoniot curve of Be.
The radius cutoff is half of the cubic cell length.
The cutoff radii of $g(r)$ are 2.37, 2.15, 1.95, and 2.00 \AA~ for 32-atom cell in
(b), (c), (d), and (e), respectively.
The cutoff radii of $g(r)$ are 4.74, 4.30, 3.90, and 4.00 \AA~
for 256-atom cell in (b), (c), (d), and (e), respectively.}
\label{fig:hug}
\end{center}
\end{figure*}

The accuracy of the TDDPMD method is demonstrated in Fig.~\ref{fig:hug}(a),
which illustrates the principal Hugoniot curves of Be as computed by both FPMD and TDDPMD methods, as well as from experiments \cite{82A-Ragan,98L-Cauble,97JAP-Nellis}.
The temperature changed from 0.4 to 2500 eV, while the density ranges from 3.50 to 8.25 g/cm$^3$.
First, we find that the FPMD results agree well with the available experimental data, demonstrating the accuracy of the FPMD method. Next,
Fig.~\ref{fig:hug}(a) shows that the two curves from FPMD and TDDPMD agree excellently,
suggesting that the fitted deep neural network
is adequate to describe the equation-of-state of warm dense Be across a wide range of temperatures and densities.

Furthermore,
%
the aforementioned drastic changes of temperatures and ion densities severely affect the local structures of Be atoms,
and we select four representative points along the principal Hugoniot curve of Be with different temperatures and densities to plot the
radial distribution functions of $g(r)$ in Figs.~\ref{fig:hug}(b)-(e).
In a general view, the resulting $g(r)$ from FPMD has a large oscillation, which is mainly caused by the limited length of trajectory from the computationally costly FPMD simulations. On the other hand, $g(r)$ from TDDPMD are more smooth since the efficiency of TDDPMD is much higher than FPMD and does not change with temperature. Therefore, more accurate structural properties are possible to obtain via TDDPMD with larger system sizes and longer trajectories.
To be specific, We find that the shape of $g(r)$ substantially changes along the Hugoniot curve.
First of all, when the system is in the conditions of 2 eV, 4.5 g/cm$^3$, and 683 GPa, we see the first two peaks in $g(r)$ correspond to two shells of local structures around Be. Due to the imposed periodic boundary conditions,
the cutoff radius of $g(r)$ can only be chosen up to half of the cell length.
Therefore, in a 32-atom cell with a density of 4.5 g/cm$^3$, the second shell structure of Be cannot be accurately obtained.
In this regard, we further adopt a 256-atom cell in TDDPMD to see the changes of shell structures of Be.
Second, when the temperature is elevated to 11 eV with a density of 6.0 g/cm$^3$ and pressure 2674 GPa,
the second peak vanishes, which is an indication that Be has only one shell of neighbors remains in this condition.
Third, the first peak disappears under the conditions of 200 eV, 8.1  g/cm$^3$, and 78358 GPa.
Finally, when the system is in the conditions of 1000 eV, 7.5 g/cm$^3$, and 395642 GPa, the $g(r)$ becomes non-zero at a shorter distance of $r$, suggesting that the averaged distance between ions is shortened at higher temperatures.
\begin{figure*}[htbp]
\begin{center}
\includegraphics[width = 16cm]{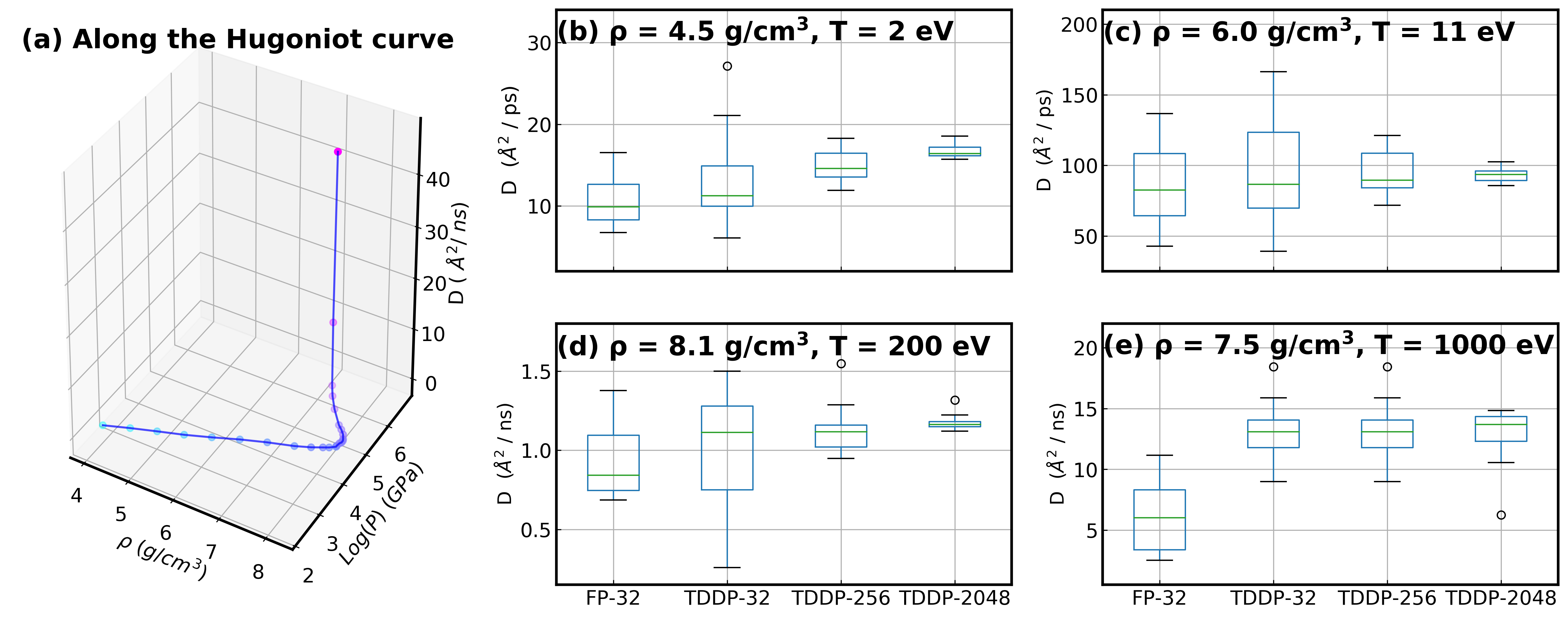}
\caption{(Color online)
(a) Diffusion coefficients $D$ of warm dense Be along the principal Hugoniot curve in terms of
different pressures $P$ (in GPa) and ion densities $\rho$ (in g/cm$^3$).
The 32-atom cell is utilized in TDDPMD simulations.
(b)-(e) Diffusion coefficients $D$ for selected ion densities $\rho$ (in g/cm$^3$) and electron temperatures $T$ (in eV). A 32-atom cell is used in FPMD
simulations while 32-, 256, and 2048-atom cells are adopted in TDDPMD simulations. We use box plot for illustration of ten diffusion coefficients $D$ at each condition, which are obtained by dividing the whole trajectory into ten parts and calculating the diffusion coefficient for each part. The upper and lower black lines denote the maximums and minimums. The blue upper and lower bounds of boxes represent quartiles of $D$, the green lines depict medians of $D$, and individual points are outliers.
}
\label{fig:d}
\end{center}
\end{figure*}

Transport properties are of particular interest in warm dense matter, and the traditional FPMD cannot estimate accurate diffusion coefficients due to the small system size and short trajectory adopted.
In this regard, the efficient TDDPMD method is capable of solving the above issues
by simulating larger systems and longer trajectories.
We test the diffusion coefficients $D$ of warm dense Be in Fig.~\ref{fig:d}.
First, the diffusion coefficients $D$ of Be are plotted along the Hugoniot curve in Fig.~\ref{fig:d}(a),
when the pressure is lower than $\sim3\times$10$^4$ GPa,
we find that the diffusion coefficients $D$ of Be steadily increases with the density.
However, a significant increase of diffusion coefficient $D$ is observed at a higher pressure above the inflection point.
%
We further investigate the size effects by selecting four systems and testing diffusion coefficients $D$ from 32-, 256-, and 2048-atom cells, the results are shown in Figs.~\ref{fig:d}(b)-(e).
In general, a larger value of diffusion coefficient $D$ is found in a larger cell and the corresponding error bars are smaller, in consistence with previous works~\cite{04JPCB-Yeh}. As a result,
these tests demonstrate the
necessity to use a larger cell to reduce the size effects and converge diffusion coefficients of warm dense matter.

In conclusion, we propose the TDDPMD method for efficient warm dense matter simulations.
The method is based on training the FPMD trajectories and yields a deep neural network to accurately describe the free energies and forces of atoms in a wide range of electron temperatures.
In particular, the TDDPMD method largely increases the fitting accuracy as compared to the DPMD method~\cite{18L-Zhang} by including the scaling factor, which is constructed based on
our observation that the logarithm of deviations of free energies and atomic forces are proportional to $log(T)$.
We demonstrated the excellent performances
of TDDPMD in reproducing the principal Hugoniot curve and associated structural and dynamical properties of warm dense Be.
Furthermore, diffusion coefficients of Be along the Hugoniot curve were obtained
with larger systems and longer trajectories. We expect that the newly proposed
TDDPMD method could have a profound impact in studying properties of materials in extreme conditions, especially for those properties that need large systems and long trajectories, such as the transport properties or phase transition.
More broadly, the temperature parameter $T$ utilized in the TDDPMD method can be generalized to other parameters that play the role of macroscopic quantities.
Finally, the current training data used in this work are generated by extensive FPMD simulations.~\cite{16PP-He}
The generation of training data would be too expensive if one targets at a much wider region of phase space or more complex configurations.
To this end, one would need an active learning procedure to generate uniformly accurate models with a minimal set of training data~\cite{zhang2019active}.
These issues will be investigated in future studies.

\bibliography{ref}

\end{document}